\documentclass{aa}

\usepackage{amssymb,amsmath,hyperref,txfonts,amsfonts}
\usepackage{xcolor}
\usepackage{mathtools}
\usepackage{mathrsfs}

\usepackage{textcomp} 
\usepackage{color}
\usepackage{mathtools}
\usepackage{xspace}
\usepackage{comment}
\usepackage{multirow}
\usepackage{url}
\usepackage{hyperref}
\usepackage{epstopdf}
\usepackage{ragged2e}
\usepackage{tabularx}
\usepackage{url}
\usepackage{soul}
\usepackage{textcomp}
\usepackage{gensymb}
\usepackage{mathrsfs}
\usepackage{needspace}
\usepackage[T1]{fontenc}
\usepackage[english]{babel}
\usepackage[utf8]{inputenc}
\usepackage{float}
\usepackage{longtable}
\usepackage[encapsulated]{CJK}
\usepackage{ucs}

\hypersetup{colorlinks,citecolor=blue,linkcolor=blue,urlcolor=blue}

\newcommand{\mw}[1]{\textcolor{black}{#1}}

\newcommand\sbullet[1][.5]{\mathbin{\vcenter{\hbox{\scalebox{#1}{$\bullet$}}}}}

\def\sgra{\object{Sgr~A*}\xspace}
\def\m87{\object{M87$^{\ast}$}\xspace}

\defcitealias{Issaoun2022}{Issaoun, Wielgus et al. 2022}

\begin{document}

   \title{The internal Faraday screen of Sagittarius A*}

   \author{Maciek Wielgus
          \inst{1,2},
          Sara Issaoun\inst{3,4},
          Iv\'an Martí-Vidal\inst{5,6},
          Razieh Emami\inst{3},
          Monika Moscibrodzka\inst{7},\\
          Christiaan D. Brinkerink\inst{7},
          Ciriaco Goddi\inst{8,9,10,11},
          \and
          Ed Fomalont\inst{12}
          }

   \institute{Max-Planck-Institut f\"ur Radioastronomie, Auf dem H\"ugel 69, D-53121 Bonn, Germany\\
         \email{maciek.wielgus@gmail.com}
                      \and
                Institute of Physics, Silesian University in Opava, Bezru\v{c}ovo n\'{a}m. 13, CZ-746 01 Opava, Czech Republic
                      \and
             Center for Astrophysics | Harvard \& Smithsonian, 60 Garden Street, Cambridge, MA 02138, USA
             \and
             NASA Hubble Fellowship Program, Einstein Fellow
         \and Departament d'Astronomia i Astrof\'{\i}sica, Universitat de Val\`encia, C. Dr. Moliner 50, E-46100 Burjassot, Val\`encia, Spain
\and Observatori Astronòmic, Universitat de Val\`encia, C. Catedr\'atico Jos\'e Beltr\'an 2, E-46980 Paterna, Val\`encia, Spain
            \and
             Department of Astrophysics, Institute for Mathematics, Astrophysics and Particle Physics (IMAPP), Radboud University, P.O. Box 9010, 6500 GL Nijmegen, The Netherlands
             \and
Instituto de Astronomia, Geof\'isica e Ci\^encias Atmosf\'ericas, Universidade de S\~ao Paulo, R. do Matão, 1226, S\~ao Paulo, SP 05508-090, Brazil
\and          
Dipartimento di Fisica, Universit\'a degli Studi di Cagliari, SP Monserrato-Sestu km 0.7, I-09042 Monserrato (CA), Italy
\and
INAF - Osservatorio Astronomico di Cagliari, via della Scienza 5, I-09047 Selargius (CA), Italy
\and
INFN, sezione di Cagliari, I-09042 Monserrato (CA), Italy
\and
             National Radio Astronomy Observatory, 520 Edgemont Road, Charlottesville, VA 22903, USA
             }

\abstract
 { We report on 85-101\,GHz light curves of the Galactic Center supermassive black hole, Sagittarius A* (\sgra), observed in April 2017 with the Atacama Large Millimeter/submillimeter Array (ALMA). This study of high-cadence full-Stokes data provides new measurements of the fractional linear polarization at a 1-2\% level resolved in 4\,s time segments, and stringent upper limits on the fractional circular polarization at 0.3\%. We compare these findings to ALMA light curves of \sgra at 212-230\,GHz observed three days later, characterizing a steep depolarization of the source at frequencies below about 150\,GHz. We obtain time-dependent rotation measure (RM) measurements, with the mean RM at 85-101\,GHz being a factor of two lower than that at 212-230\,GHz. Together with the rapid temporal variability of the RM and its different statistical characteristics in both frequency bands, these results indicate that the Faraday screen in \sgra is largely of internal character, with about half of the Faraday rotation taking place inside the inner 10 gravitational radii, contrary to the common external Faraday screen assumption. We then demonstrate how this observation can be reconciled with theoretical models of radiatively inefficient accretion flows for a~reasonable set of physical parameters. Comparisons with numerical general relativistic magnetohydrodynamic simulations suggest that the innermost part of the accretion flow in \sgra is much less variable than what these models predict, in particular, the observed magnetic field structure appears to be coherent and persistent.}

   
   \titlerunning{The internal Faraday screen of Sagittarius A*}
   \authorrunning{Wielgus et al.}

\maketitle

\section{Introduction}

Sagittarius A* (\sgra) is the radio source associated with a 4$\times$10$^6 M_{\odot}$ supermassive black hole (SMBH) located in our Galactic Center \citep{Do2019,Gravity2022,SgraP1}. 
The source is characterized by a particularly low mass accretion rate of $\sim 10^{-8} M_{\odot}{\rm yr}^{-1}$ \citep{Quatertaet2000,Yuan2003,SgraP5}. Spectral energy distribution (SED) analysis allowed to identify it as an advection dominated / radiatively inefficient type of accretion flow \citep[ADAF/RIAF; ][]{Narayan1995,Yuan2014,SgraP2,SgraP5}. The SED of \sgra exhibits a maximum at a turnover frequency in the range of several hundred GHz that can be attributed to synchrotron emission from the hot thermal electrons in the marginally optically thin innermost region of the accretion disk \citep{Yuan2003}. Despite significant progress on the theoretical front and multiple observational studies across the electromagnetic spectrum since \sgra was first identified in 1974 \citep{Balick1974}, our detailed understanding of this object and the accretion flow surrounding it remains incomplete \citep[for a recent review see, e.g., ][]{Morris2023}.

\begin{figure*}[t]
    \centering
    \includegraphics[width = 0.49\textwidth]{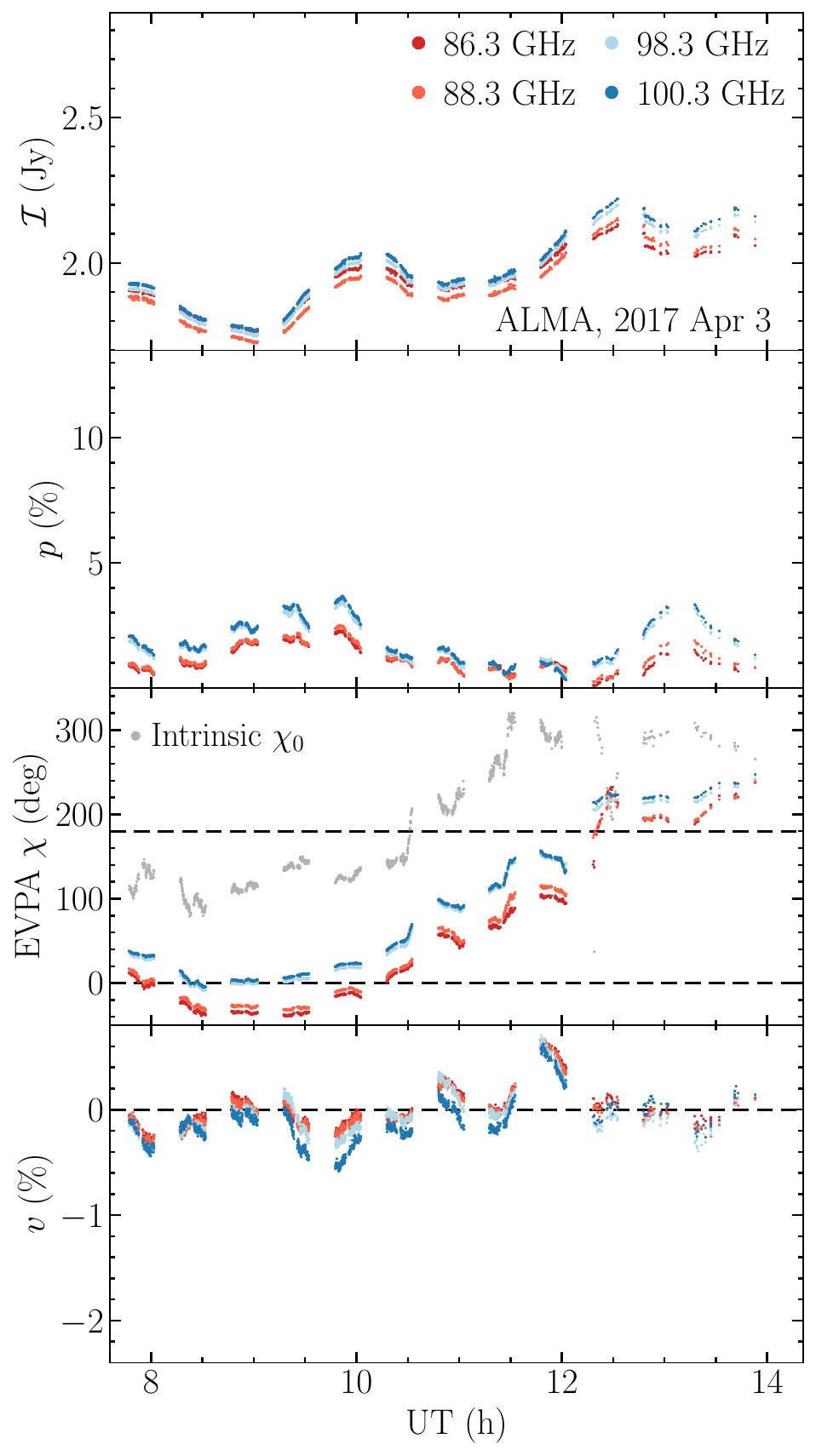}
    \includegraphics[width = 0.49\textwidth]{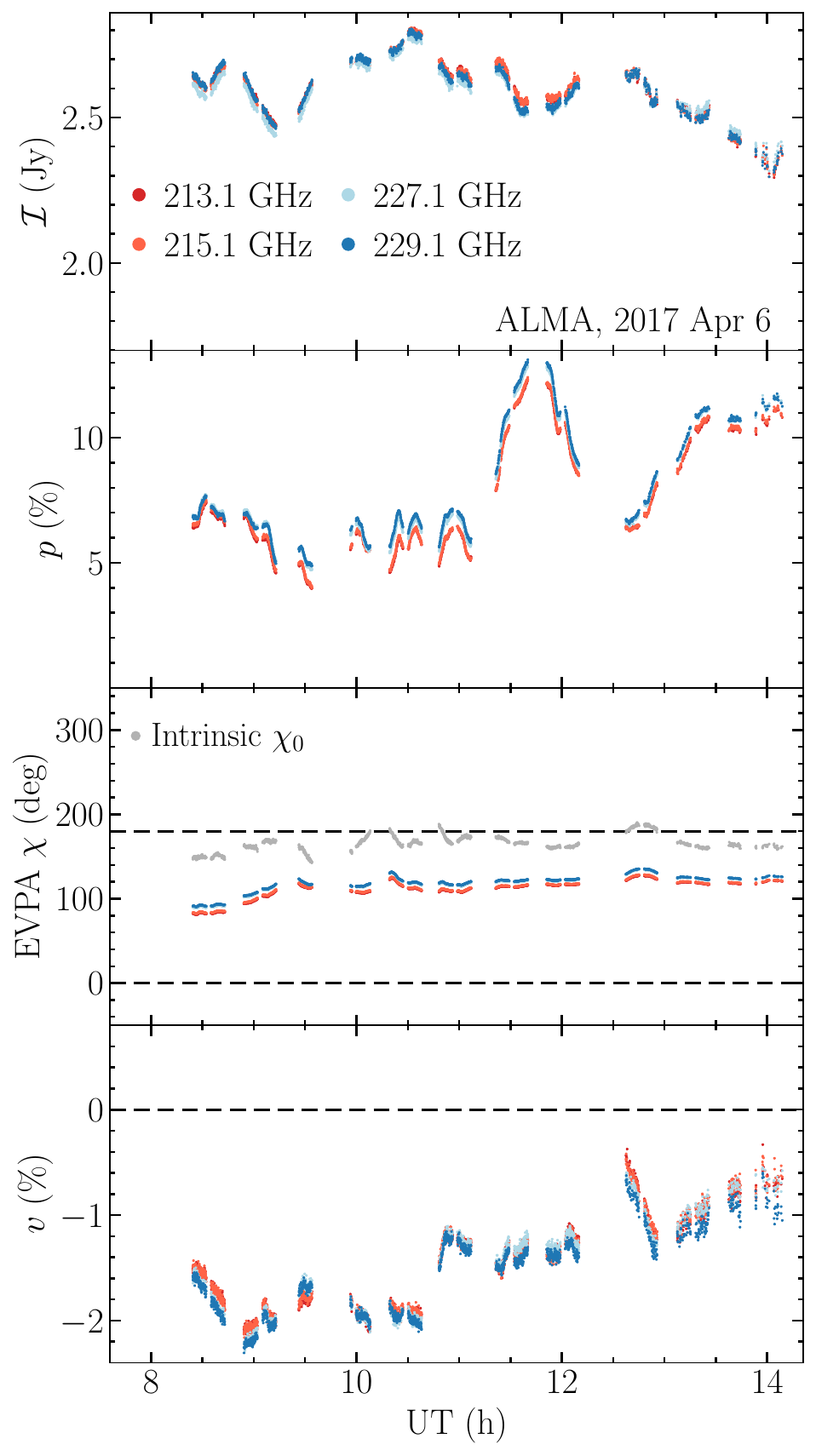}
    \caption{Overview of the ALMA light curves of \sgra obtained in band 3 (left column) and band 6 (right column), each with four 2\,GHz-wide sub-bands, spanning 85-101\,GHz and 212-230\,GHz, respectively. A significant reduction of the fractional polarization and an increase of the EVPA variability in the lower frequency band are visible.
}
    \label{fig:light_curves}
\end{figure*}

Very long baseline interferometric (VLBI) radio observations indicate that the intrinsic size of \sgra decreases with frequency, reaching event horizon scales at $\sim$230\,GHz \citep[e.g.,][]{Doeleman2008,SgraP1}. At $\sim$90\,GHz the intrinsic diameter of the VLBI image corresponds to about 20 gravitational radii $r_{\rm g} = GM_{\sbullet[1.35]}/c^2$ \citep{Shen2005, Bower2006,Lu2011,Issaoun2019}, hence the emission originates in large part from a layer external to the innermost scales $\lesssim 5 r_{\rm g}$. Therefore, observations at different radio frequencies can be used to probe distinct regions of the accretion flow. Linear (LP) and circular (CP) polarization, rotation measure (RM), as well as temporal variability of these observed quantities, provide additional constraints on \sgra models \citep[e.g.,][]{Quatertaet2000,Bower2003,Marrone2007, Sharma2007, Pang2011, Ressler2023}.

The CP of \sgra was first reported by \citet{Bower1999_CP} at 4.8\,GHz, while the first LP detections were obtained by \citet{Aitken2000} in the 150-400\,GHz range. Multiple subsequent observations showed LP values of $\sim$5-10\% at around 230\,GHz \citep{Bower2003,Marrone2007,Bower2018,Wielgus2022_LP}. The source becomes strongly depolarized at lower frequencies, and there is a very limited number of LP detections below 150\,GHz in the literature \citep{Bower1999_LP,Macquart2006,Liu2016}. Measurements of weak negative CP were reported at frequencies below 20\,GHz \citep{Bower1999_CP,Bower2002}. Above 200\,GHz a stronger CP $\sim -1\%$ appears \citep{Munoz2012,Bower2018,Wielgus2022_LP}. Only weak upper limits were known for intermediate frequencies so far \citep{Tsuboi2003}.

In this paper we present a study of ALMA light curves in the 85-101\,GHz range (ALMA band 3), including unambiguous high-time-cadence detections of variable LP, stringent upper limits on CP, and time-resolved measurements of Faraday rotation. We compare these results to the study of 212-230\,GHz (ALMA band 6) light curves obtained in a quasi-contemporaneous epoch \citep{Wielgus2022_LC,Wielgus2022_LP}. At both frequency bands we build on the results of \citet{Goddi2021}, where a preliminary analysis of \sgra polarization in the same ALMA observations, reduced under an unphysical static source assumption, was presented. The main consequence of our results for the \sgra system is the observationally-demonstrated presence of a significant internal component of the Faraday screen, which must be located within the central 10\,$r_{\rm g}$ region (Section~\ref{sec:observations}). We discuss the implications of this measurement for the RIAF accretion flow model, utilizing VLBI observations to constrain the radial distribution of the electron temperature (Section~\ref{sec:interpretation}). We also compare the measurements with predictions from general relativistic magnetohydrodynamic (GRMHD) simulations (Section~\ref{sec:grmhd}). A brief summary is given in Section~\ref{sec:summary}.

\section{Observations and data analysis}
\label{sec:observations}

\begin{table*}[th!]
     \caption{Summary of the \sgra light curves observed with ALMA in April 2017
     }
    \begin{center}
    \setlength{\tabcolsep}{3.4pt}
    \renewcommand{\arraystretch}{1.1} 
    \begin{tabularx}{0.985\linewidth}{lcccccccccc}
    \hline
    \hline
      & \multicolumn{4}{c}{ Band 3, 2017 Apr 3}  &  &\multicolumn{4}{c}{ Band 6, 2017 Apr 6}   \\
\hline
      $\nu_{\rm obs}$ (GHz) & 86.3 & 88.3 & 98.3 & 100.3 & & 213.1 & 215.1 & 227.1$^a$ & 229.1\\
      $\mathcal{I}$ (Jy) & $1.91\pm 0.09$ & $1.87 \pm 0.09$ & $1.91\pm 0.10$ & $1.93\pm 0.10$ & &  $2.62 \pm 0.09$ & $2.62 \pm 0.09$ &  $2.62 \pm 0.09$ &  $2.61 \pm 0.09$ \\
      $|\mathcal{P}|$ (mJy) & $23.2\pm 9.5$ & $23.9\pm10.3$ & $33.4\pm 15.2$ & $36.3\pm 16.7$ & & $194.1\pm 55.5$ & $195.4\pm 55.6$ & $206.9\pm 55.5$ & $208.3\pm 55.1$ \\
      $p$ (\%) & $1.23\pm 0.51$ & $1.28 \pm 0.56$ & $1.79\pm 0.81$ & $1.89\pm 0.87$ & &
      $7.45\pm 2.26$ & $7.50\pm 2.25$ & $7.93\pm 2.26$ & $8.03 \pm 2.28$ \\
      $\chi^b$ (deg) & $-19.4 \pm 44.1$ & $-15.1 \pm 41.3$ & $8.7\pm 37.6$ & $13.6\pm 38.5$ & &
      $109.9\pm 11.7$ & $110.8\pm 11.6$ & $116.4\pm 11.2$ & $117.3\pm 10.9$ \\
      $\mathcal{V}$ (mJy) & $0.4\pm 4.1$ & $0.1 \pm 4.0$ & $-0.6\pm 4.4$ & $-2.3\pm 4.6$ & &
      $-39.4\pm 11.1$ & $-39.3\pm 11.1$ & $-39.9 \pm 11.5$ & $-41.2 \pm 10.7$ \\
      $v$ (\%)$^c$ & $0.0 \pm 0.2$ & $0.0 \pm 0.2$ & $0.0\pm 0.2$ & $-0.1 \pm 0.2$ & &  
      $-1.50 \pm 0.41$ & $-1.50 \pm 0.40$ & $-1.52 \pm 0.42$ & $-1.57 \pm 0.38$ \\
       $\alpha_{\mathcal{I}}$ & \multicolumn{4}{c}{ $0.10 \pm 0.01$} & &  \multicolumn{4}{c}{ $-0.01 \pm 0.01$} \\
      $\alpha_{p}$ & \multicolumn{4}{c}{ $2.97 \pm 0.06$} & &  \multicolumn{4}{c}{ $0.91 \pm 0.02$} \\
      $\alpha_{v}$ & \multicolumn{4}{c}{ --- } & &  \multicolumn{4}{c}{ $0.50 \pm 0.02$} \\
      RM$^d$ & \multicolumn{4}{c}{ $-2.14 \pm 0.51 $} & & \multicolumn{4}{c}{ $-5.04 \pm 0.83 $} \\
     \hline    
    \hline
    \end{tabularx}
    \label{tab:LC_statistics}
    \end{center}
    $^a$scaled up by 4\% to account for the CN absorption line \citep[Appendix H.1. of][]{Goddi2021};\
    $^b$calculated using directional statistics; \\
    $^c$band 3 values are consistent with nondetection, upper limit of $|v| < 0.3\%$; \ $^d$ in the units of $10^5$ rad m$^{-2}$
    
\end{table*}

\begin{figure*}[h]
    \centering
    \includegraphics[height=1.9in,trim=-0.0cm -0.0cm 0cm 0cm,clip]{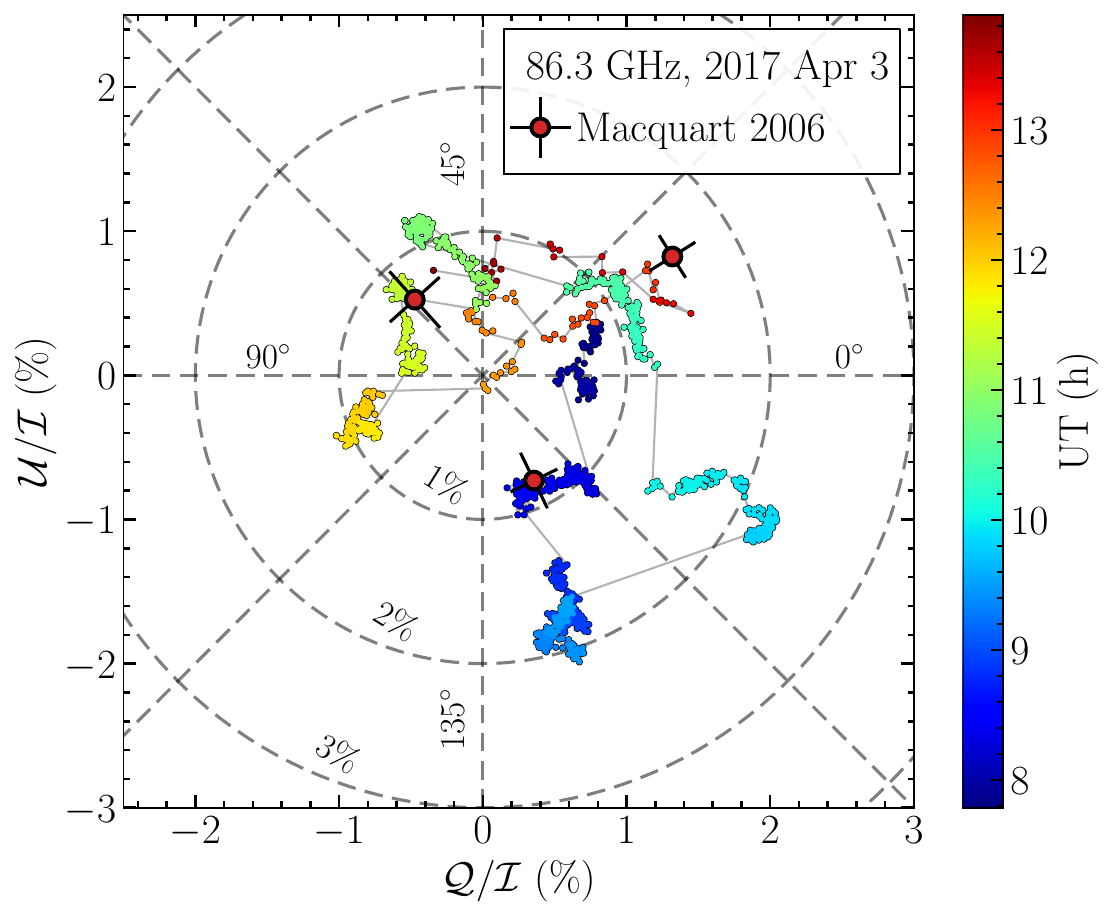}
    \includegraphics[height=1.9in,trim=-0.0cm -0.0cm 0cm 0cm,clip]{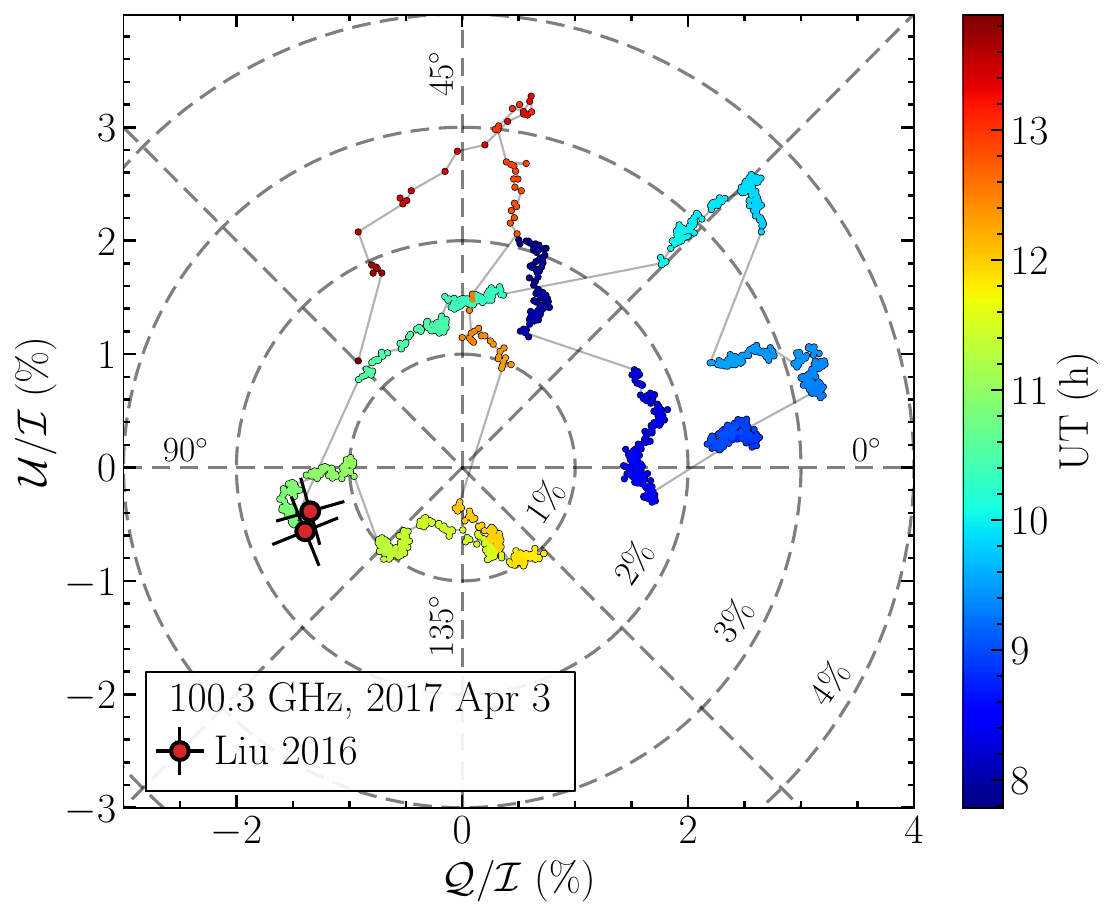}
     \includegraphics[height=1.9in,trim=-0.0cm -0.0cm 0cm 0cm,clip]{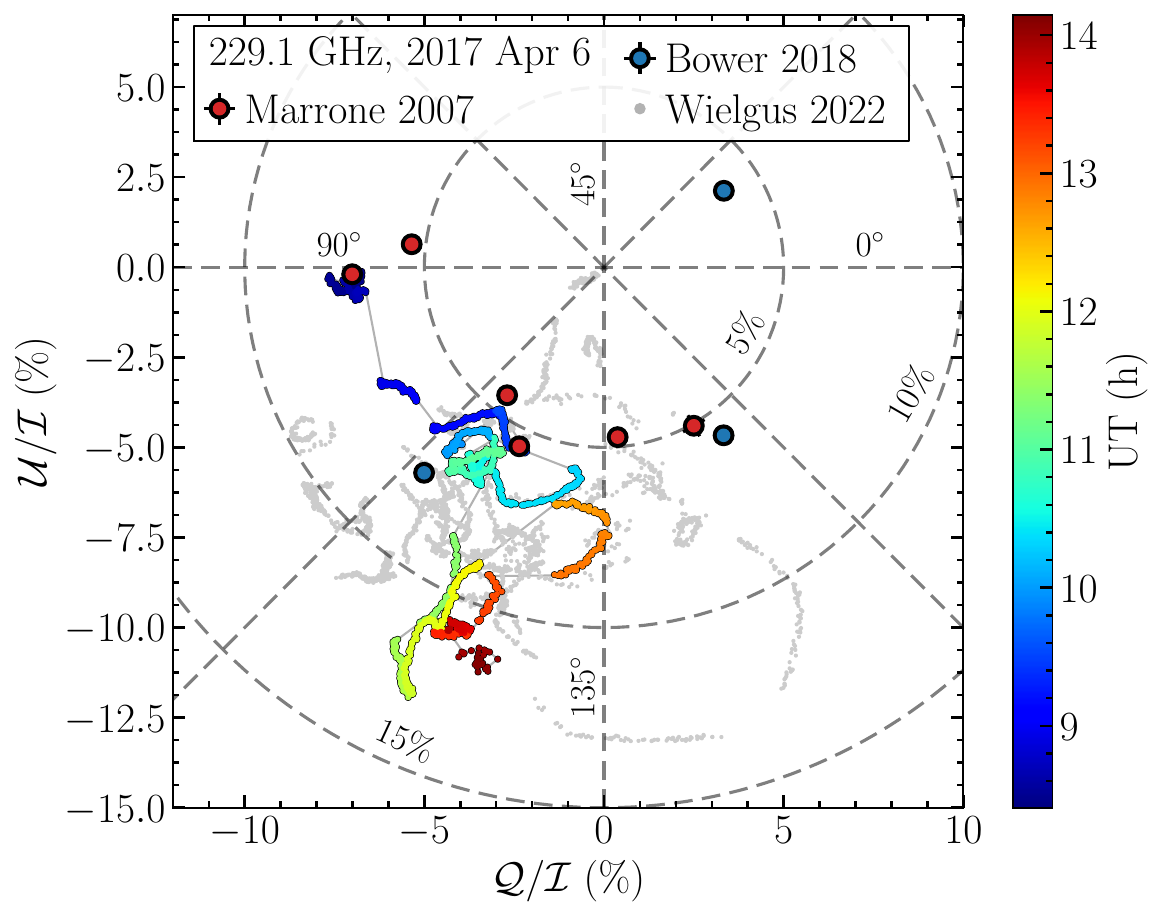}
    \caption{Time-dependent fractional LP measurements obtained at 86.3, 100.3, and 229.1\,GHz presented on the $\mathcal{Q}/\mathcal{I}-\mathcal{U}/\mathcal{I}$ plane of linear polarization, compared with the past measurements at the corresponding frequencies. Colors denote the time progression, following the colorbars. Polar coordinates indicate fractional polarization (in \%) and EVPA (in degrees; $0^\circ \equiv 180^\circ$ corresponds to EVPA aligned with the North-South axis).
}
    \label{fig:QUplots}
\end{figure*}

\subsection{Data reduction and conventions}

The high sensitivity of ALMA enabled detailed recent studies of the rapid variability of \sgra \citep{Iwata2020,Murchikova2021,Wielgus2022_LC}. In April 2017, \sgra was observed by ALMA during its participation in the VLBI campaigns with the Global mm-VLBI Array \citep[GMVA;][]{Issaoun2019,Goddi2019} and with the Event Horizon Telescope \citep[EHT;][]{SgraP1, Goddi2021} as a compact phased array \citep{Matthews2018,Goddi2019}. Algorithms enabling the extraction of ALMA-only time-dependent light curves of the compact \sgra source from these data were developed and extensively discussed in \citet{Wielgus2022_LC}. The method is based on the well-motivated assumption of negligible variation of the arc-minute scale radio emission surrounding \sgra \citep{Lo1983,Mus2022}, allowing to self-calibrate to the extended source structure and  extract the time-dependent point source component, reducing the impact of fluctuating amplitude gains. This robust approach, employed following the ALMA QA2 calibration \citep{Goddi2019}, has been applied to band 6 observations at 212-230\,GHz, obtained on 2017 April 6, 7, and 11, and presented in \citet{Wielgus2022_LC,Wielgus2022_LP}. In this paper, we employ a very similar calibration algorithm to the data obtained on 2017 April 3 in ALMA band 3, with 4 frequency sub-bands, each 2\,GHz wide, centered at 86.3, 88.3, 98.3, and 100.3\,GHz. The sub-band depolarization and decorrelation effects are negligible in all cases, hence we work with the data averaged in sub-bands.
The only difference with respect to the procedure of \citet{Wielgus2022_LC} was motivated by the wide total fractional bandwidth in band 3 observations $\sim$15\%, as opposed to $\sim$7\% in band 6. In order to account for the frequency dependence of the extended source structure across a wider frequency coverage we tested two approaches: carrying out an independent CLEAN deconvolution of each sub-band and using a~CLEAN multi-frequency-synthesis first-order expansion \citep{Conway1990,Casa2007}. Both variants of the extended emission modeling resulted in very consistent light curves of the compact component, fitted individually in each sub-band. The latter approach was adopted for the final data set presented in this paper.

The data set contains all Stokes parameters, that is, total intensity $\mathcal{I}$, linear polarization $\mathcal{P} = \mathcal{Q} + i\mathcal{U}$, and circular polarization $\mathcal{V}$, observed with a time cadence of 4\,s. We define the fractional LP $p$ and fractional CP $v$ as
\begin{equation}
    p = \frac{|\mathcal{P}|}{\mathcal{I}} = \frac{\sqrt{\mathcal{Q}^2 + \mathcal{U}^2}}{\mathcal{I}}   \ \ ; \ \ v = \frac{\mathcal{V}}{{\mathcal{I}}} \ .
\end{equation}
Furthermore, we define the electric vector position angle (EVPA) as $\chi = 0.5 {\rm Arg}(\mathcal{Q} + i\mathcal{U})$.

\subsection{Intensity and polarization}

\begin{figure}[h]
    \centering
    \includegraphics[height=2.1in,trim=-0.0cm -0.0cm 0cm 0cm,clip]{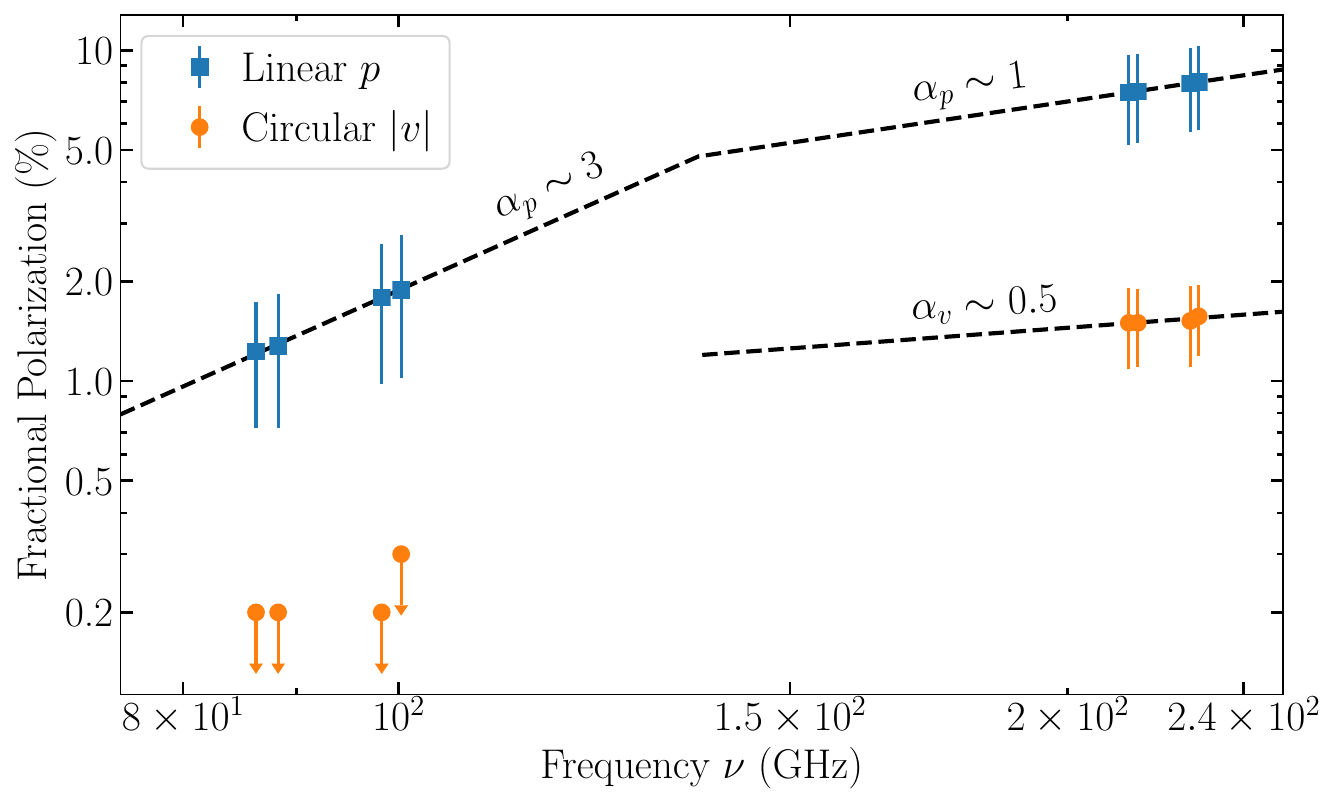}

    \caption{Spectral dependence of \sgra fractional polarization, fitted with a power law $p(\nu) \propto \nu^\alpha$. Both LP and CP measurements indicate a rapid depolarization with decreasing frequency at about 100\,GHz (2017 Apr 3), in contrast to a weak dependence of fractional polarization at observing frequencies above 200\,GHz (2017 Apr 6). Error bars shown represent the standard deviation in the samples and are dominated by the intrinsic source variability, while the formal uncertainties of the spectral indices fits are very small, see Tab. \ref{tab:LC_statistics}. 
}
    \label{fig:spectra_index}
\end{figure}

The obtained \sgra light curves are summarized in Fig.~\ref{fig:light_curves}, with mean values and standard deviations reported in Tab.~\ref{tab:LC_statistics}. The reported uncertainties are strongly dominated by the intrinsic source variability, with the effective signal-to-noise ratio (S/N) of each 4\,s ALMA measurement $\sim$200 for the Stokes $\mathcal{I}$ component \citep{Wielgus2022_LC}. We focus on comparisons between band 3 observations on 2017 April 3 and band 6 observations on 2017 April 6, given that these light curves are closest in time and of similar total duration. Generally similar mean parameters were found on other days of the band 6 observations \citep[see Appendix~\ref{app:table_711} and][]{Wielgus2022_LC,Wielgus2022_LP}, with some discrepancies highlighted in Appendix~\ref{app:table_711}. We employed an identical flagging procedure for each data set presented in Fig.~\ref{fig:light_curves}, in order to remove points for which the calibration procedures did not converge. The sparser time coverage after $\sim$12:30\,UT reflects the decreasing elevation of \sgra at ALMA by the end of the observing epoch. We verified that flagging these data does not appreciably impact the reported results. Our physically-motivated analysis (Figs.~\ref{fig:light_curves}-\ref{fig:spectra_index}) confirms that there is a remarkable change in the degree of polarization between the two ALMA bands, first reported under a~static source assumption by \citet{Goddi2021}. Furthermore, the EVPA is significantly more variable at 85-101\,GHz than at 212-230\,GHz on the same timescales (Figs.~\ref{fig:light_curves}-\ref{fig:QUplots}). 

\begin{figure*}[t]
    \centering
    \includegraphics[width=3.13in,trim=-0.0cm -0.60cm 0.2cm 0cm,clip]{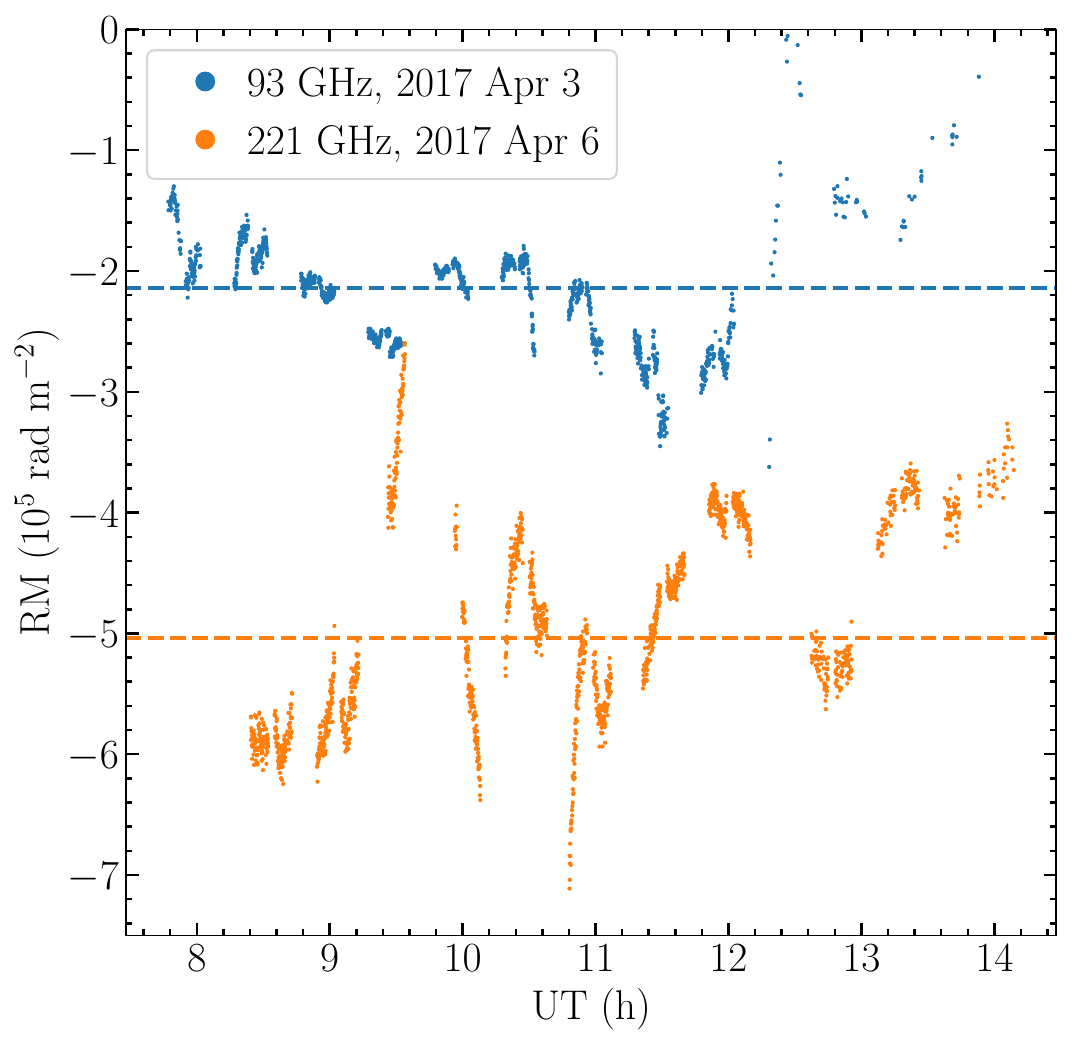}
    \includegraphics[width=1.355in,trim=0.2cm 0.1cm 0cm 0cm,clip]{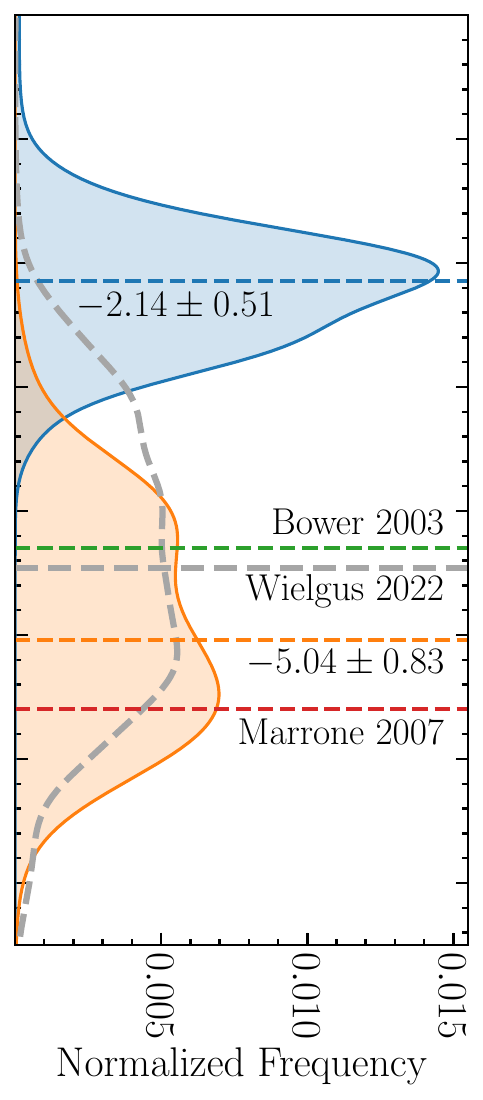}
\includegraphics[width=2.385in,trim=-0.20cm -0.8cm 0cm 0cm,clip]{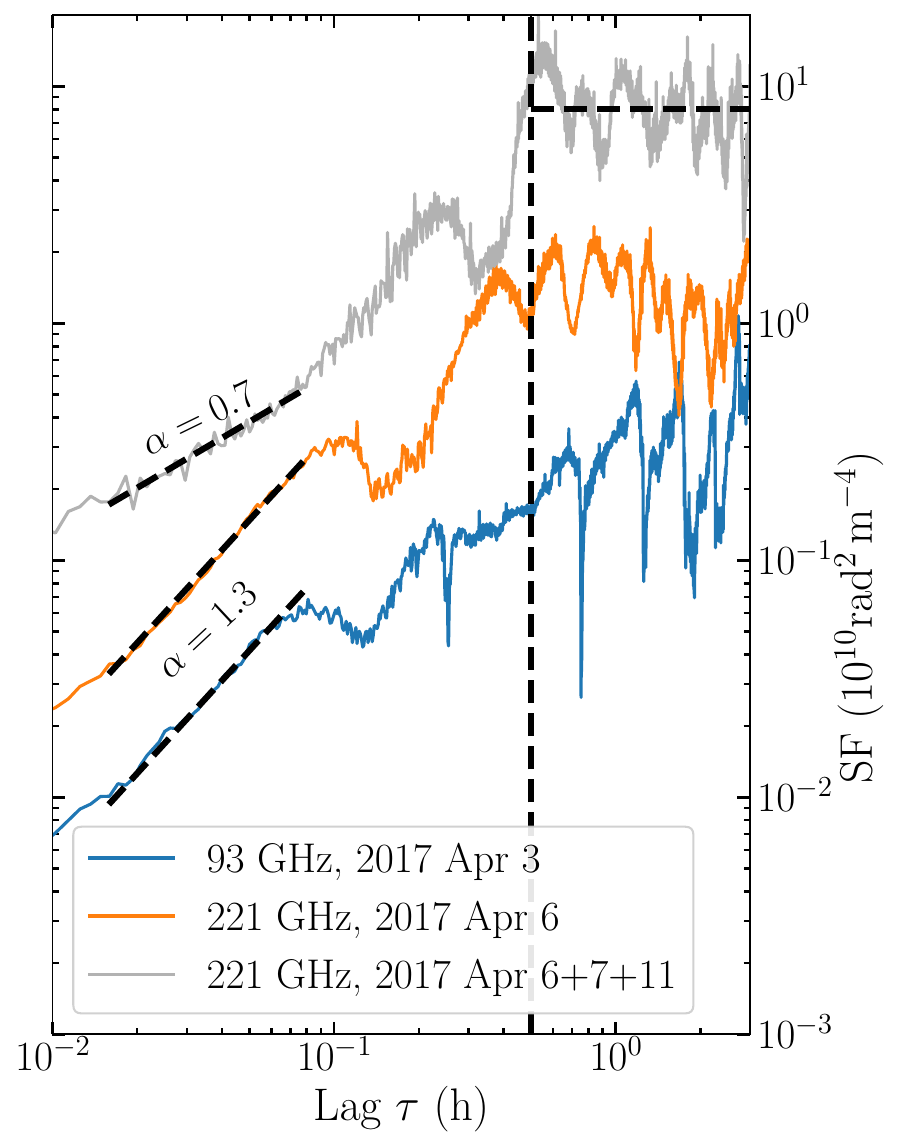}
    \caption{\textit{Left center:} Time-resolved RM measurements in ALMA band 3 (85-101\,GHz, blue) and band 6 (212-230\,GHz, orange) in April 2017, and corresponding histograms of the observed RM. Dashed lines represent mean values. The rapid variability of RM hints at the internal character of the Faraday screen, which is further supported by a significant discrepancy between the measurements obtained for the two bands, also including historical data, and a joint histogram of the 2017 Apr 6, 7, and 11 data (dashed gray line). \textit{Right:} SF analysis of the RM variability, indicating more variability at larger frequencies, with relatively more variation on the shortest timescales. A 0.5\,h variability decorrelation timescale is found for the 221\,GHz RM data, which is absent at 93\,GHz.
    }
    \label{fig:RMplot}
\end{figure*}

We also present time-dependent fractional LP vector measurements obtained at 86.3, 100.3, and 229.1\,GHz \citep[the latter following][]{Wielgus2022_LP} in Fig. \ref{fig:QUplots}. We compare these measurements with the results reported by \citet{Macquart2006} at 86.3\,GHz (low S/N measurements not shown), \citet{Liu2016} at 105-107\,GHz, \citet{Marrone2007} at 230.6-231.9\,GHz, \citet{Bower2018} at 226.0\,GHz, and \citet{Wielgus2022_LP} at 229.1\,GHz. We observe consistency between these historical measurements and our light curves at 86.3 and 100.3\,GHz. The large temporal variations of the EVPA that we observe on a few hours timescale explain EVPA changes reported by \citet{Macquart2006} between subsequent observing days. These variations appear as a full counter-clockwise loop on the plane of linear polarization executed between 8:00 and 11:30 UT on 2017 Apr 3, see the EVPA panel of Fig.~\ref{fig:light_curves} and the first two panels of Fig.~\ref{fig:QUplots}. If we attempted to explain this loop in the framework of a coherent orbital motion, we would conclude Keplerian orbit of $\sim$20$r_{\rm g}$ radius. However, the direction of the loop's direction is opposite to the one observed by \citet{Wielgus2022_LP} and \citet{gravity2023}, challenging that interpretation. At 230\,GHz, we generally see a slightly increased degree of LP in 2017 with respect to the previous measurements, which can be related to the lower total compact flux density reported in April 2017 \citep[mean 2.4~Jy in April 2017 comparing to mean 3.2~Jy in 2005-2019;][]{Wielgus2022_LC}. The 230\,GHz EVPA, while variable, appears to have a preference for values $\chi \sim 120^\circ$ (derotated EVPA $\chi_0 = \chi - {\rm RM}\lambda^2 \sim 170^\circ$), and almost never takes values in the range between $-20^\circ$ and $80^\circ$, which constitutes a~potentially powerful constraint on theoretical models, particularly for the on-sky position angle of the system. As an example, for a toy model in which the observed non-zero net LP is a consequence of Doppler boosting of the approaching side of the inclined accretion disk, we would expect the on-sky projected disk spin axis to align with the intrinsic EVPA $\chi_0$ for the predominantly azimuthal magnetic field, and to be perpendicular to $\chi_0$ if the magnetic field is predominantly vertical. 

\subsection{Depolarization at lower frequencies}
\label{sect:depolar}

A significant change of the fractional polarization with the observing frequency is summarized in Fig.~\ref{fig:spectra_index}, where we fitted spectral indices of $p$ and $|v|$, as reported in Tab. \ref{tab:LC_statistics}, across ALMA sub-bands. In band 3, we measured a steep depolarization toward lower frequencies $p \propto \nu^3$, inconsistent with the milder depolarization $p \propto \nu$ seen in band 6. Similarly, fractional CP has a~weak dependence on the frequency in band 6, inconsistent with the upper limits on $|v|$ that we report in band 3, which we give as $|v| < 0.3\%$, based on the time-resolved analysis. An increase of LP fraction with frequency has been observed in active galactic nuclei (AGN), e.g., \citet{Agudo2014} reported a mean factor of $\sim$1.5 change between 86 and 229\,GHz, corresponding to a fractional LP spectral index $\alpha_p \approx 0.5$. A~plausible explanation is that the higher frequency emission originates from a~more compact region, with a more ordered magnetic field, or that the changes are related to the optical depth variation. However, the effect observed in \sgra is far more extreme, hinting at a~prominent transition in the system in its innermost part, at around $5-10\,r_{\rm g}$, possibly separating strong and ordered magnetic fields near the event horizon \citep{Johnson2015} from a weaker and more chaotic component further away. The depolarization could also be caused by the optical depth $\tau$ increasing at lower frequencies, as polarization from the optically thick thermal synchrotron emission is suppressed exponentially with $\tau$ \citep{Pacholczyk1970}. Hence, an increase in the optical depth by $\Delta \tau \lesssim 2$ would suffice to explain the change in fractional polarization. However, the opacity interpretation is only straightforward under the assumption of a uniform optical depth in the emission zone, which is likely an oversimplification.

\subsection{Faraday rotation}
\label{sect:RM}

Linearly polarized radiation undergoes a change in EVPA as it propagates through a Faraday screen -- the magnetized plasma located between the emitter and the observer. This effect is quantified with rotation measure (RM), which can be defined as
\begin{equation}
    {\rm RM} = \frac{{\rm d} \chi }{ {\rm d} \lambda^2 } \approx \frac{\Delta \chi}{ \Delta \lambda^2} = \frac{ \chi(\lambda_2)- \chi(\lambda_1) }{ \lambda^2_2 - \lambda^2_1 }
    \label{eq:RM_basic}
\end{equation}
for two observing wavelengths $\lambda_1$, $\lambda_2$. If the Faraday depth is small and the Faraday screen is external with respect to the emitting region, the RM is independent of the observing wavelength and the approximation in Eq.~\ref{eq:RM_basic} turns into a strict equality. The intrinsic (derotated) EVPA of the emission may then be calculated as
\begin{equation}
\chi_0 = \chi(\lambda) - {\rm RM} \lambda^2 \ .
\label{eq:intrinsicEVPA}
\end{equation}
On the other hand, measuring inconsistent values of RM at different wavelengths indicate a deviation from the $\lambda^2$ relation, implying either a complex unresolved source structure involving multiple Faraday screens, or an internal Faraday screen overlapping with the emission zone at some wavelengths \citep{Burn1966,Brentjens2005}. Such effects have been observed, as an example, in the quasar  3C\,273 \citep{Hovatta2019}. In the case of \sgra, VLBI observations reveal a persistent, simple, compact, single-component source morphology across frequencies, hence the internal Faraday screen interpretation is favored.

The RM toward \sgra is well established through measurements at frequencies near 230\,GHz ($\lambda = c/\nu \approx 1.3$\,mm), with ${\rm RM} \approx -5 \times 10^5$\,rad m$^{-2}$ \citep{Bower2003,Marrone2007,Bower2018,Wielgus2022_LP}, which is a rather large value when compared to what is typically observed toward AGN sources \citep[e.g.,][]{Goddi2021}. The RM was shown to fluctuate significantly \citep{Bower2018}, even on very short sub-hour timescales \citep{Wielgus2022_LP}. This poses a difficulty in estimating the RM based on non-simultaneous EVPA measurements, with both RM and EVPAs fluctuating in time, and the short associated timescales generally point toward the Faraday screen compactness.

There are not many RM measurements in \sgra available at different frequencies. They involve estimates at 345\,GHz by \citet{Marrone2007}, who found a mean RM in excess of $-10^6$\,rad m$^{-2}$, but large measurement uncertainties did not allow to exclude consistency with the values observed at 230\,GHz. At longer wavelengths, \citet{Macquart2006} calculated RM between 86\,GHz and 230\,GHz, but concluded that consistency with 230\,GHz measurements depended on shifting the (180$^\circ$-periodic) EVPA measurement by 180$^\circ$, and significantly lower RM was estimated without the shift. A preliminary discussion of ALMA bands 3 and 6 RM measurements (under the static source assumption) and their implications was presented in \citet{Goddi2021}.

Following our time-dependent calibration of the ALMA observations, we estimated RM as a function of time by fitting a~linear model for the EVPA as a function of squared wavelength across 4 sub-bands in band 3\mw{, independently for each timestamp}. Subsequently, we compared these measurements with the band 6 results, already discussed in Appendix A of \citet{Wielgus2022_LP}. While the time-dependent model linear in $\lambda^2$ provides a good fit quality across each ALMA band individually, the resulting RM measurements significantly differ between the two bands. This is shown in Fig.~\ref{fig:RMplot}, where high time cadence RM measurements in $\sim$6\,h duration observing windows are presented for the mean band 3 frequency of 93\,GHz (2017 April 3) and mean band 6 frequency of 221\,GHz (2017 April 6). RM values observed at the lower frequency band are incompatible with both quasi-contemporaneous and historical measurements near 230\,GHz, demonstrating a deviation from the $\lambda^2$ relation defined in Eq.~\ref{eq:RM_basic}. This implies that about half of the Faraday rotation occurs internally with respect to the 93\,GHz emission region, that is, at most several $r_{\rm g}$ away from the SMBH's event horizon. The two measurements give an approximate scaling relation ${\rm RM} \propto r^{-1}$.

 Furthermore, in the EVPA panels of Fig.~\ref{fig:light_curves} we show the intrinsic EVPA values $\chi_0(t)$, derotated using the time-dependent RM measurements and Eq.~\ref{eq:intrinsicEVPA}. If the Faraday screen was external, but variable on short timescales, derotating $\chi(t)$ should reduce its variability. Since the shape of derotated $\chi_0(t)$ and its measured variability remain overall very similar to those of the observed $\chi(t)$, we conclude that the intrinsic variability of the emitter dominates over the variability in the external Faraday screen, consistent with the dominant character of the internal Faraday screen component.

Simultaneous measurements of LP at distinct frequencies are necessary in order to provide an ultimate and bulletproof argument for the $\lambda^2$ relation violation, otherwise one could still attempt to explain our results with an unusually low, but $\lambda$-independent RM on 2017 April 3. Nonetheless, these findings constitute by far the most convincing observational demonstration of an internal Faraday screen component in \sgra to date.

\subsection{Quantifying the RM variability}

We investigate the RM variability with the structure function (SF) approach \citep{Simonetti1985}, which is the time-domain analog of the power spectrum analysis, defined as 
\begin{equation}
    {\rm SF}(\tau) = \langle ( {\rm RM}(t) - {\rm RM}(t- \tau) )^2 \rangle \ ,
\end{equation}
where the averaging is taken over all times $t$. Hence, the SF effectively splits the observed variation across timescales $\tau$ (lags). The results of the SF analysis are shown in the right panel of Fig.~\ref{fig:RMplot}, where we compare 2017 Apr 3 and 6 analysis, as well as the joint analysis of band 6 RM calculated for Apr 6, 7, and 11, with a total of $\sim$ 20\,h of data. We confirm significantly less variation in the lower frequency band across the sampled timescales in the absolute sense. Relative to the mean value, RM fluctuates by $\sim$24\% at the lower frequency and by $\sim$16\% at the higher one, Tab.~\ref{tab:LC_statistics}. Furthermore, the short timescale variability slope in band 3 is significantly steeper than the one revealed by the joint analysis of band 6, implying less power at the shortest timescales in band 3 data, consistent with the Faraday rotation occurring on larger physical scales at the lower frequency. We observe a similar dependence in the SF of RM in a~numerical GRMHD simulation of \sgra (see also Section~\ref{sec:grmhd}), with the lower frequency slope $\alpha = 0.5$ steeper than the higher frequency one $\alpha = 0.2$. However, both slopes in the simulation are significantly less steep than the observations, indicating a relatively larger contribution from variability occurring on the shortest timescales in the numerical model than in the real source. In the joint analysis of the Apr 6, 7,11 data, an SF maximum occurs at around 0.2-0.3\,h, and for timescales longer than 0.5\,h SF becomes flat, indicating uncorrelated variability structure for larger time lags. Hence, we identify 0.5\,h as a characteristic timescale for the 221\,GHz Faraday screen variability. A similar flattening is not apparent at 93\,GHz, but the reason is most likely related to the short total duration of the observed light curve. We predict that a longer decorrelation timescale will be identified in band 3 with additional observations, allowing to contrast the characteristic timescales at different frequencies. The frequency dependence of the statistical characteristics of the RM variability is yet another hint that the Faraday screen is different for the two observed bands, and at least partly cospatial with the compact emission region.

\section{Implications for the RIAF model}
\label{sec:interpretation}

\subsection{Electrons temperature profile}
\label{sec:estimate_Te}

\begin{table*}[th!]
     \caption{VLBI measurements of the intrinsic morphology and brightness temperature of \sgra in April 2017.
     }
    \begin{center}
    \setlength{\tabcolsep}{3.4pt}
    \renewcommand{\arraystretch}{1.15} 
    \begin{tabularx}{0.965\linewidth}{lccccccccc}
    \hline
    \hline
    Reference & Date (2017) & $\nu$ (GHz) & $S_\nu$ (Jy) & $\theta_{\rm maj}$ (mas) & $\theta_{\rm maj}$ (mas) & $r_{\rm e}\,(r_{\rm g})$ & $T_{\rm B, obs}$ ($10^{10}$ K) & $(1+z)$ &  $T_{\rm B, int}$ ($10^{10}$ K) \\

    \citet{Cho2022} & 3 Apr & 22.2 & 1.0 & 0.80 & 0.60 & 66.9 & 0.51 & 1.02 & 0.52 \\
    \citet{Cho2022} & 4 Apr & 43.1 & 1.3 & 0.30 & 0.23 & 24.8 & 1.24 & 1.04 & 1.29 \\
    \citet{Issaoun2019} & 3 Apr & 86.3 & 1.9 & 0.10 & 0.12 & 9.7 & 2.58 & 1.12 & 2.89 \\
    \citet{SgraP1,SgraP3} & 6-7 Apr & 228.1 & 2.4 & 0.052 & 0.052 & 4.1 & 2.08 & 1.40 & 2.91 \\
     \hline

     \hline
    \hline
    \end{tabularx}
    \label{tab:TB_VLBI}
    \end{center}
\end{table*}

In order to study the implications of our findings for the RIAF model \citep{Yuan2003}, we first use the quasi-simultaneous VLBI observations of \sgra from April 2017 to constrain the radial distribution of the electron temperature. These data sets are summarized in Tab. \ref{tab:TB_VLBI}. The VLBI observations measure flux density $S_{\nu}$ and allow to estimate the intrinsic (descattered, see \citealt{Johnson2018}) angular size of the source $\theta_{\rm maj} \times \theta_{\rm min}$, modeled as an elliptical Gaussian. Hence, they allow to estimate the brightness temperature of \sgra, a~proxy for the~temperature of the emitting electrons under the assumption of a~thermal energy distribution
\begin{equation}
    T_{\rm B, obs} = 1.22 \times 10^{12} \frac{S_\nu}{\nu^2 \theta_{\rm maj} \theta_{\rm min}} \ \ {\rm [K]} \ ,
\end{equation}
where flux density is given in Jy, frequency in GHz, and angular dimensions in mas. 
\begin{figure}[h]
    \centering
    \includegraphics[width=3.4in,trim=-0.0cm -0.0cm 0cm 0cm,clip]{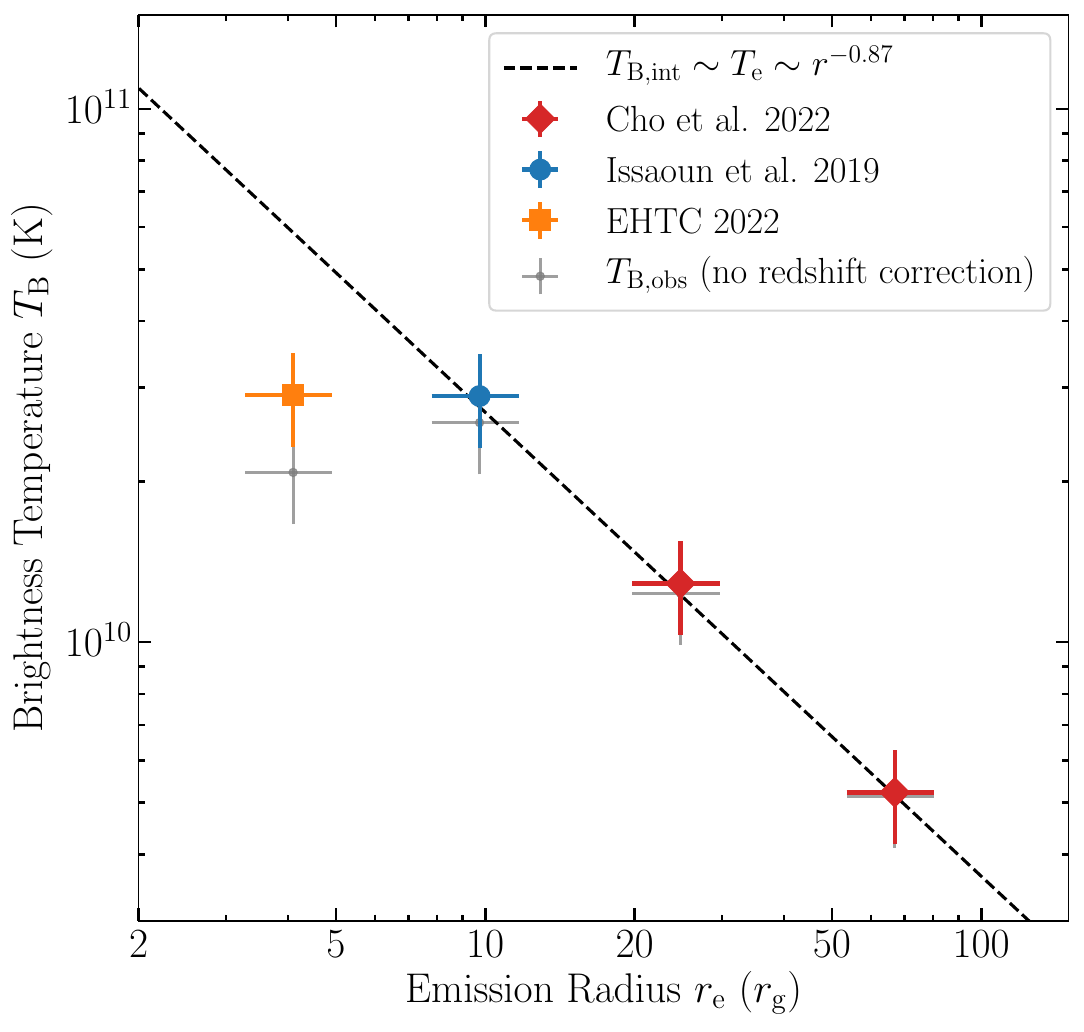}
    \caption{Fitting intrinsic brightness temperature $T_{\rm B,int}$ as a function of the emission radius $r_{\rm e}$. $T_{\rm B, int}$ is a proxy for the temperature of the emitting electrons $T_{\rm e}$. The measurements are based on the gravitational-redshift-corrected VLBI observations performed on 2017 Apr 3-7, with 20\% errorbars assumed on both $T_{\rm B}$ and $r_{\rm e}$ estimates. 
}
    \label{fig:fit_Te}
\end{figure}
Furthermore, the observed intrinsic size of the source allows to estimate the characteristic emission radius $r_{\rm e}$ as
\begin{equation}
    r_{\rm e}/r_{\rm g} = 0.5(\theta_{\rm maj} \theta_{\rm min})^{0.5}/\theta_{\rm g} - 1 \ ,
    \label{eq:emission_radius}
\end{equation}
where $\theta_{\rm g} = r_{\rm g}/D$ is the angular size of the source gravitational radius $r_{\rm g}$ viewed from a distance $D$, and following \citet{Gravity2022} we take $\theta_{\rm g} = 5.1\, \mu$as. The subtraction of $1$ in Eq.~\ref{eq:emission_radius} approximately accounts for the geometric lensing around the black hole regardless of the detailed spacetime geometry \citep{Gralla2020,Wielgus2021}. The emission from the vicinity of the SMBH's event horizon is affected by the gravitational redshift, which scales the 
intrinsic brightness temperature by a factor of $(1+z) = [-g_{ tt}(r_{\rm e})]^{-0.5} = (1-2M/r_{\rm e})^{-0.5}$, $T_{\rm B, int} = (1+z) T_{\rm B, obs}$, see Fig.~\ref{fig:fit_Te} and Tab.~\ref{tab:TB_VLBI}. 

The EHT measurement deviates from a power law characterising $T_{\rm B, int}$ at lower frequencies, which can be clearly seen in Fig.~\ref{fig:fit_Te}. We verify that the choice of elliptical Gaussian model rather than a ring model with parameters reported in \citet{SgraP1} does not impact the $T_{\rm b}$ estimates by more than $\sim 20\%$. A~possible explanation for the lower $T_{\rm B, int}$ is either that the emission at 230\,GHz, corresponding to $r_{\rm e} \approx 4 r_{\rm g}$, is optically thin and non-thermal, violating assumptions of the $T_{\rm b}$ calculation, and/or that the emission is produced by plasma plunging into the black hole with a relativistic radial velocity component and thus experiencing Doppler deboosting. Hence, we fit a power law $T_{\rm e}(r) = T_{0} (r/r_{\rm g})^\gamma$ only to the three 22-86\,GHz data points, obtaining $\gamma = -0.87 \pm 0.20$ and $T_{0} \approx 2.0 \times 10^{11}$ K. The estimated power law index is consistent with the values commonly assumed for $T_{\rm e}(r)$ in RIAF models \citep[e.g.,][]{Broderick2011,Vincent2022,Vos2022} based on spectral energy density fitting \citep{Yuan2003}. Additionally, based on the four VLBI measurements reported in Tab.~\ref{tab:TB_VLBI}, we model the intrinsic source size dependence on frequency, finding a scaling of $r_{\rm e} \propto \nu^{-1.2}$, in between the values reported by \citet{Shen2005} and \citet{Bower2006}.

\subsection{RM in a RIAF model}

We study the properties of a simple spherically symmetric RIAF power law model to see whether the observed wavelength-dependent RM can be reconciled with the theoretical expectations. In this framework, we can compute RM with the integral along the line of sight from the emitter to the observer \citep[e.g.,][]{Moscibrodzka2017}
\begin{equation}
    {\rm RM}= 10^4 \frac{e^3}{ 2 \pi m_{\rm e}^2 c^4 } \int\limits^{\rm r_{\rm obs}}_{r_{\rm e}} n_{\rm e} B_{\parallel} f_{\rm e}(\Theta_{\rm e}) \text{d}r \ [{\rm rad \,m}^{-2}] \ ,
    \label{eq:RM}
\end{equation}
for the electron number density $n_{\rm e}$ measured in cm$^{-3}$ and magnetic field component along the line of sight $B_\parallel$ measured in G. The additional dimensionless multiplier $f_{\rm e}(\Theta_{\rm e})$ is related to the reduced impact of hot relativistic electrons, and can be expressed as a~function of dimensionless electron temperature $\Theta_{\rm e} = kT_{\rm e}/ m_{\rm e} c^2$ \citep{Quatertaet2000,Ressler2023}
\begin{equation}
f_{\rm e}(\Theta_{\rm e}) = \begin{cases}
1 &\text{if $\Theta_{\rm e}  \le 1$} \ ,\\
\Theta_{\rm e}^{-2} \left[ 0.5 \log(\Theta_{\rm e}) \left( 1 - \Theta_{\rm e}^{-1} \right)  + 1\right] &\text{if $\Theta_{\rm e} > 1$ \ .}
\end{cases}
\end{equation}
The presence of this correction factor was raised as an argument against the internal Faraday screen in a RIAF system, since the contribution from the very hot innermost region would be strongly reduced \citep[e.g.,][]{Marrone2006,Macquart2006}. The argument appears reasonable, as the suppression factor from $f_{\rm e}(\Theta_{\rm e})$ reaches $0.008$ at $T_{\rm e} = 10^{11}$K ($\Theta_{\rm e} = 16.8$). Moreover, the characteristic linear scale of a compact system is small in comparison to what the thickness of the external Faraday screen could be. Nevertheless, we will show that constructing a RIAF flow model with significant internal Faraday rotation occurring inside the $10 r_{\rm g}$ radius is feasible regardless of the $f_{\rm e}(\Theta_{\rm e})$ factor.

\begin{figure}[h]
    \centering
    \includegraphics[width=3.4in,trim=-0.0cm -0.0cm 0cm 0cm,clip]{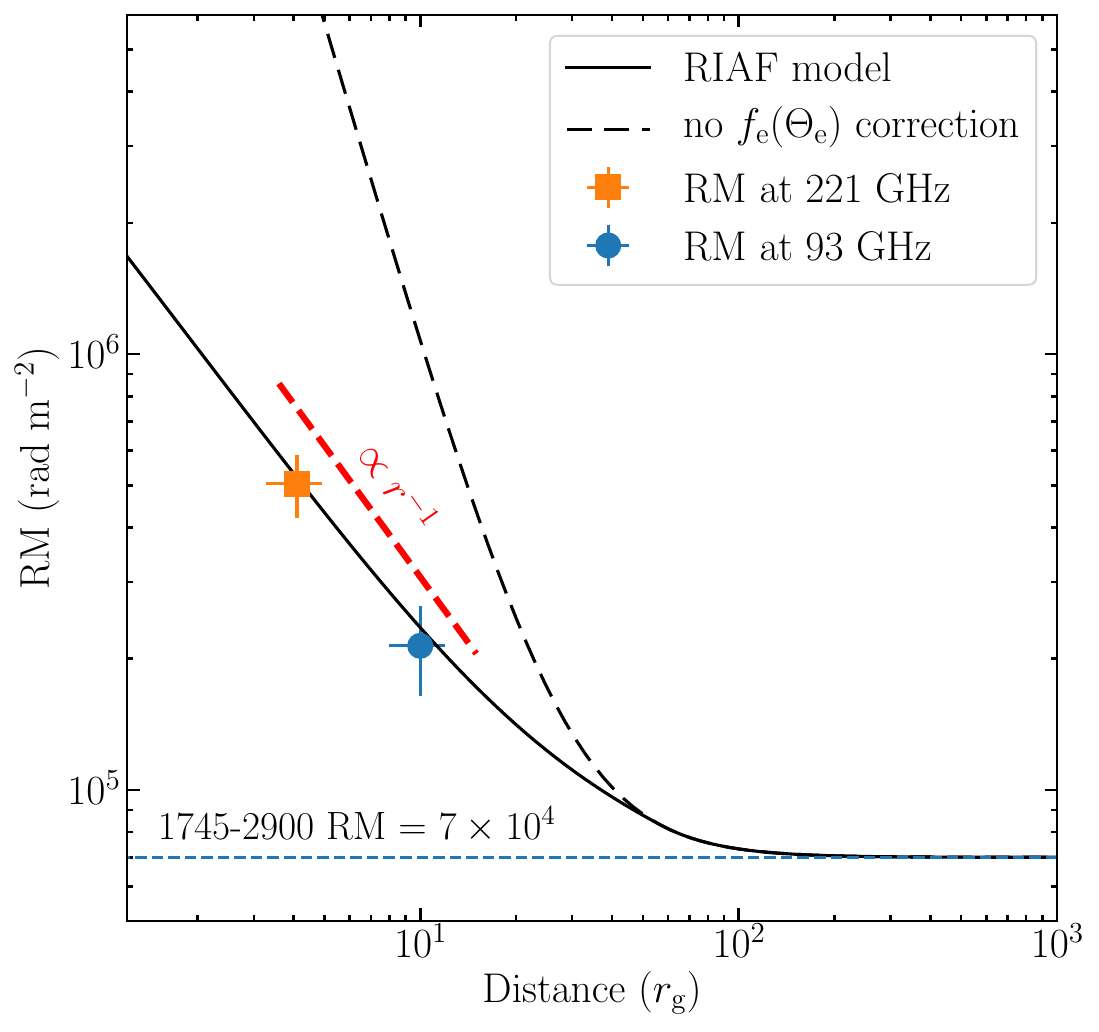}
    \caption{Predicted value of RM as a function of the emission radius for a RIAF toy-model fitted to the ALMA data. The background RM corresponding to the value observed toward the magnetar J1745-2900 is assumed. Accounting for the relativistic correction $f_{\rm e}$ decreases the RM significantly, by a factor of 20 for $r_{\rm e}= 4 r_{\rm g}$, but does not preclude the dominant contribution to the Faraday rotation from the innermost part of the flow.
}
    \label{fig:riaf_RM}
\end{figure}

For the RIAF model we assume the electron temperature distribution estimated in Section \ref{sec:estimate_Te}
\begin{equation}
    T_{\rm e}(r) = 2 \times 10^{11} \left(\frac{r}{r_{\rm g}} \right)^{-0.87}  {\rm [K]} \ .
\end{equation}
We also assume power law distributions for the electron number density $n_{\rm e}$ and magnetic field parallel to the line of sight $B_{\parallel}$
\begin{align}
    & n_{\rm e}(r) = n_0 \left(\frac{r}{r_{\rm g}} \right)^{\alpha}  {\rm [cm}^{-3}] \ , \\
    & B_{\parallel}(r) = B_0 \left(\frac{r}{r_{\rm g}} \right)^{\beta}  {\rm [G]} \ .
\end{align}
To simplify the problem further, we assume the emission at each separate frequency band to originate from a sphere located at the radius $r_{\rm e}$, as estimated in Subsection \ref{sec:estimate_Te} -- an onion-like model of the \sgra radio source. Particularly in case of a low optical depth these radii may only have an effective, approximate sense. We then request for the RM, integrated between the respective $r_{\rm e}$ and the distant observer ($r_{\rm obs} = 10^5 r_{\rm g}$ for practical purposes) using Eq.~\ref{eq:RM}, to match the ALMA measurements. We assume a background RM value of $-7 \times 10^4$ rad m$^{-2}$, following the measurement of RM toward the Galactic Center magnetar J1745-2900 \citep{Eatough2013}, located $\sim10^6 r_{\rm g}$ away from \sgra (projected distance of $\sim$0.1 pc), with a caveat that the RM toward the magnetar is itself variable, and may involve a contribution from its compact vicinity. Since the sign of RM depends only on the polarity of the uniform magnetic field, we work with absolute values of the RM. With two RM measurements corresponding to different $r_{\rm e}$, we can then fit for the two model parameters, $\alpha + \beta$, and $n_0 B_0$. In Fig.~\ref{fig:riaf_RM} we show a~solution, corresponding to
\begin{align}
& \alpha + \beta = -3.5  \ , \\
& \left(\frac{n_0}{1 \times 10^7 {\rm cm}^{-3}} \right) \left(\frac{B_0}{500 G} \right) = 1 \ .
\end{align}
Comparing our results to the thermal synchrotron one-zone emission model given by \citet{SgraP5}, with $n_{\rm e, zone} = 10^6$\,cm$^{-3}$ and $B_{\rm zone} = 29$\,G, we can reproduce these numbers reasonably well as $n_{\rm e}(r_{\rm e})$ and $B_\parallel(r_{\rm e})$ for $\alpha = -1.5$, $\beta = -2$, and $r_{\rm e} = 4 r_{\rm g}$. In Fig.~\ref{fig:riaf_RM} we also show the RM from the same model, computed neglecting the $f_{\rm e}(\Theta_{\rm e})$ factor. Clearly, accounting for this correction has a big impact on the results, but regardless of its presence Faraday rotation may be dominated by the plasma in the innermost part of the accretion flow.

To reproduce the observed ratio of the two RM measurements, around a factor of 2, we generally need a high value of $\alpha+\beta$ in the RIAF model. As an example $\alpha + \beta = -2.5$ assumed by \citet{Vos2022} appears insufficient, yielding a ratio of 1.4. Steeper radial decay favors the interpretation involving vertical, rather than azimuthal, magnetic fields, but on the other hand it may also correspond to a partially inhomogeneous magnetic field, where contributions from different emission regions cancel one another. If we attribute the large contribution to the Faraday rotation from the innermost region of the flow to the partially ordered \citep{Johnson2015} vertical magnetic field then it also suggests a relatively low viewing angle, so that a significant component of the magnetic field is oriented along the line of sight. Both vertical magnetic field and low inclination are consistent with the conclusions of \citet{gravity_loops_2018}, \citet{Wielgus2022_LP}, and \citet{SgraP5}. Low viewing angle and dominant RM contribution from compact scales were also concluded by \citet{Sharma2007} based on the analysis of global MHD simulations of accretion in \sgra.

As an additional sanity check we may estimate the locations of photospheres $r_{\rm ph}$ for the observing frequencies in the model with the parameters estimated above, following calculations of \citet{Mahadevan1996} and integrating the resulting thermal synchrotron opacities inwards. Interestingly, we obtain $r_{\rm ph}$ around 4$r_{\rm g}$ at 228.1\,GHz and $r_{\rm ph}$ around 7$r_{\rm g}$ at 86.3\,GHz, which are not terribly inconsistent with the estimated emission radii $r_{\rm e}$. At lower frequencies $r_{\rm ph}$ are too low and generally smaller than $20 r_{\rm g}$. We do not attempt to fit the locations of photospheres by tuning the model parameters.

Since the presented model constitutes an extreme simplification of reality, our findings should be considered primarily as a~demonstration that an internal Faraday screen in \sgra, dominated by the contribution from very compact scales, is to be expected for a reasonable set of physical parameters in a RIAF system. Any more quantitative conclusions should be taken with a~sizeable grain of salt. 

\section{Comparisons to GRMHD}
\label{sec:grmhd}

Models more physically self-consistent than a power law RIAF are obtained through numerical GRMHD simulations. RM in GRMHD simulations was studied, among others, by \citet{Moscibrodzka2017} and \citet{Ricarte2020} in the context of M\,87* and recently by \citet{Ressler2023} for \sgra. \citet{SgraP5} considered a large library of GRMHD models in order to identify ones that fulfil observational constraints. Here we investigate one of the models favored by the EHT analysis (the "best-bet" model), corresponding to a magnetically arrested disk \citep[MAD;][]{Narayan2003} system, viewed at low inclination of 30 deg. High magnetization and small viewing angle are also supported by the analysis of \citealt{gravity_loops_2018} and \citealt{Wielgus2022_LP}. The model considered is also characterized by a~moderate SMBH spin $a_* = 0.5$ and an ion-to-electron temperature ratio parametrized by $R_{\rm high} = 160$ \citep[relatively cold accretion disk electrons;][]{Moscibrodzka2016,SgraP5}. While the model passed the majority of observational total intensity constraints, it has been reported \citep{SgraP5} that much like other GRMHD MAD models it significantly overproduces the variability in total intensity light curves when compared to observations \citep{Wielgus2022_LC}. The selected GRMHD simulation was performed using the KHARMA code \citep[a~GPU-enabled extension of the \texttt{iharm3D} code; ][]{Prather2021} and consecutively ray-traced in a curved Kerr spacetime using \texttt{ipole} \citep{Moscibrodzka2018}, see \citet{Wong2022} for details of the simulation generation pipeline. During ray-tracing a~scaling of the plasma density was selected in order to match the 2.4\,Jy total compact flux density observed at 228\,GHz in April 2017 \citep{Wielgus2022_LC}. Ray-traced images were then averaged over the entire field of view for each Stokes component at every time step to obtain full-Stokes simulated light curves.

In order to study the polarimetric properties of the simulated GRMHD light curves, we performed ray-tracing at four frequencies, 86.3, 100.3, 213.1, and 229.1\,GHz, to be able to mimic the the RM measurements at ALMA band 3 and band 6. Each light curve corresponds to the same GRMHD output of 1000 snapshots with a cadence of 5 $r_{\rm g}/c \approx 100$\,s. Since the light curves observed on 2017 Apr 3 and 6 have a duration of about $1000\,r_{\rm g}/c$ (about 6\,h), we effectively analyze five different realizations of the theoretical GRMHD model.

The GRMHD light curves from the selected best-bet model generally do not match the  observed polarimetric properties of \sgra. The modeled fractional LP is $p = 4 \pm 2$\% across frequencies, with no sign of depolarization in the lower band. Compared to the \sgra values presented in Tab.~\ref{tab:LC_statistics}, the model's polarization is too low in band 6, and too high in band 3. The EVPA in the simulation wanders much more than it does in the band 6 data. 
The model fractional CP is strongly variable and while the model mean values at 213-229\,GHz are around -1\%, similarly to the observed quantities, there is about seven times more variability in the simulation than is observed. Hence, the modeled CP sign is flipping on a~$\sim$1\,h timescale, while it appears to never change in the observed light curves \citep[e.g.,][]{Bower2002, Munoz2012, Wielgus2022_LP}.

\begin{figure}[h]
    \centering
    \includegraphics[width=3.4in,trim=-0.0cm -0.0cm 0cm 0cm,clip]{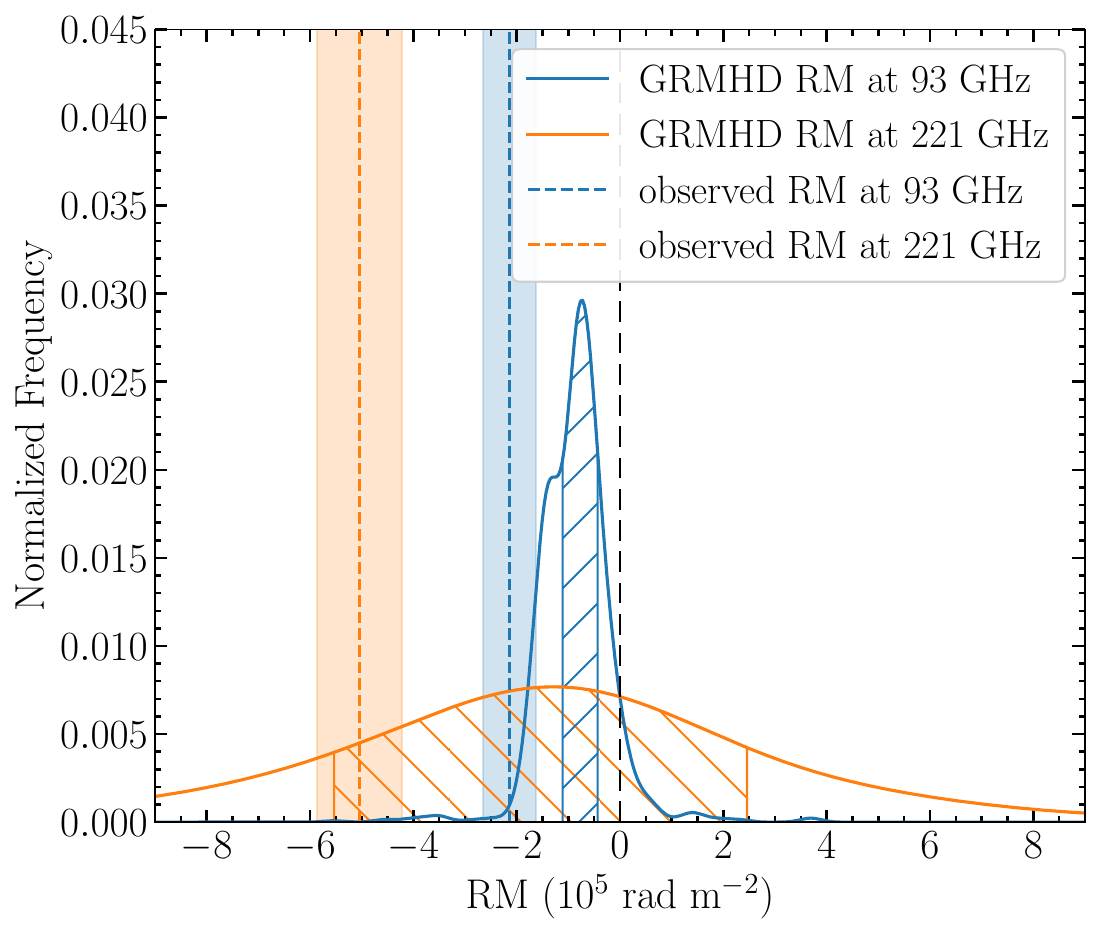}
    \caption{Smoothed histograms of RM extracted from the GRMHD simulation (sign-flipped and shifted by the Galactic Center magnetar RM $= -7\times 10^4$\,rad\,m$^{-2}$) compared to the ranges inferred from observations on 2017 Apr 3 and 6, discussed in Section \ref{sect:RM} (vertical bands). Shaded and hatched regions denote intervals containing 68\% of the distribution around the median value.
}
    \label{fig:GRMHD_RM}
\end{figure}

The RM in the simulation is evaluated from the EVPA differences between 86.3 and 100.3 GHz, as well as between 213.1 and 229.1 GHz, following Eq.~\ref{eq:RM_basic}. We only ray-trace the innermost $300\,r_{\rm g}$ zone of the GRMHD simulation, hence there is no large scale Faraday screen present. Since we find significantly different values of RM at each band for the same time snapshots, the Faraday screen in the simulations is necessarily largely internal (located between the 93 and 221\,GHz emission region), consistently with our interpretation of the observations. However, while the observed RMs vary by $\sim20-30$\% around the median values and (almost) never change sign, in the simulations the RM sign flips often, on a~$\sim$1\,h timescale. In order to maximize the consistency, in Fig.~\ref{fig:GRMHD_RM} we flip the sign of the model RM and shift the RM values by the Galactic Center magnetar RM, to roughly represent the contribution from the Faraday rotation occurring outside of the GRMHD simulation domain. Accounting for the external Faraday rotation is sufficient for the model band 3 RM to mostly remain negative, although they are about a factor of two lower than the observed values. Recently \citet{Ressler2023} suggested, based on multiscale MHD/GRMHD modeling of the \sgra accretion flow fed by stellar winds, that the constant observed RM sign is related to a large and stable external contribution, dominating over the rapidly varying RM component from the compact region. This interpretation is challenged by our results. While it is easy to imagine a larger RM bias related to an external Faraday screen, allowing to match the 93\,GHz observations, a~more challenging endeavour is to reproduce the large and strictly negative difference between RM at 221 and 93\,GHz, necessarily caused by an intrinsic Faraday screen component. Not only are the RM values in our GRMHD simulation at band 6 smaller (in a mean sense) than observed, they are also dramatically more variable with respect to the median value, which is close to zero in the simulation, see also \citet{Sharma2007} and \citet{Pang2011}. This discrepancy can be understood in the framework of the magnetic field variability, which is the only signed quantity contributing to the RM. It appears that the magnetic field is far more turbulent and variable in the simulations than in reality, which echoes the total intensity light curve variability discrepancy of \citet{SgraP5}.

MAD models with lower $R_{\rm high}$ parameter ($R_{\rm high}=160$ is the largest value considered in the EHT simulation library) will have hotter electrons than the best-bet model studied here. Hotter electrons would only further decrease the mean value of RM with little impact on its stability because the emission would still emerge from the disk with turbulent magnetic fields. An interesting alternative is the observed mm radiation originating predominantly in the jet sheath region, which is typically threaded by a more stable, nearly-vertical magnetic field in GRMHD simulations \citep{Moscibrodzka2013}. A~subsequent detailed comparison of the simulations from the EHT library (and beyond it) with polarimetric observations at different wavelengths should provide more insight in the future.

\section{Summary and conclusions}\label{sec:summary}

Using high sensitivity ALMA observations, we characterize the full-Stokes light curves of \sgra in the 85-101\,GHz range, and compare them with a complementary data set obtained at 212-230\,GHz. Both data sets were obtained just three days apart, on 2017 April 3 and 6, respectively. We provide new measurements of linear polarization at 85-101\,GHz as well as stringent upper limits on circular polarization. The fractional polarization of \sgra decreases rapidly below 150\,GHz, which we interpret as a transition in the accretion flow, possibly in magnetic field geometry, strength, or coherence, occurring at around $5-10\,r_{\rm g}$, but it could also be related to the transition to an optically thin flow at higher frequencies. Our observations yield time-dependent measurements of the RM in the 85-101\,GHz band, which we find to be lower than the established measurements at higher frequencies by a~factor of two. Together with the rapid temporal variability of the RM, lack of variability reduction in the derotated EVPA, and different statistical characteristics of RM temporal variability in the two frequency bands, these results show that the Faraday screen in \sgra is most likely largely of internal character, cospatial with the compact region of the synchrotron emission. We demonstrate how these findings can be reproduced using a simple theoretical model of a~radiatively inefficient accretion flow. Finally, we demonstrate that the Faraday screen is largely internal in the numerical GRMHD simulations of Sgr~A$^*$. However, the particular simulation that we considered, while mostly consistent with the observational total intensity constraints, did not quantitatively reproduce the observed parameters of polarization, and indicated significantly more variability in polarization fractions and in rotation measure than what is observed in \sgra.


\section*{Acknowledgements}
{We thank Geoff Bower, Ilje Cho, Michael D. Johnson, Thomas Krichbaum, Dan Marrone, Ue-Li Pen, Venkatessh Ramakrishnan, Sean Ressler, Angelo Ricarte, Pablo Torne, Sebastiano von Fellenberg, and Guang-Yao Zhao for their helpful comments and enlightening discussions, as well as Vedant Dhruv, Charles Gammie, Abhishek Joshi, Ben Prather, and George Wong for performing the GRMHD simulation that we analyzed. We also thank Alexandra Elbakyan for her contributions to the open science initiative. This paper makes use of the following ALMA data: ADS/JAO.ALMA\#2016.1.01404.V and ADS/JAO.ALMA\#2016.1.00413.V ALMA is a partnership of ESO (representing its member states), NSF (USA) and NINS (Japan),together with NRC (Canada), NSC and ASIAA (Taiwan), and KASI (Republic of Korea), in cooperation with the Republic of Chile. The Joint ALMA Observatory is operated by ESO, AUI/NRAO and NAOJ. This research is supported by the European Research Council advanced grant “M2FINDERS - Mapping Magnetic Fields with INterferometry Down to Event hoRizon Scales” (Grant No. 101018682). SI is supported by Hubble Fellowship grant HST-HF2-51482.001-A awarded by the Space Telescope Science Institute, which is operated by the Association of Universities for Research in Astronomy, Inc., for NASA, under contract NAS5-26555. IMV acknowledges partial support from Generalitat Valenciana (GenT Project CIDEGENT/2018/021), the MICINN Research Project PID2019-108995GB-C22 and the ASTROVIVES FEDER infrastructure IDIFEDER-2021-086. MM acknowledges support by the NWO grant No. OCENW.KLEIN.113 and support by the NWO Science Athena Award.
 RE acknowledges the support from grant numbers 21-atp21-0077, NSF AST-1816420, and HST-GO-16173.001-A as well as the Institute for Theory and Computation at the Center for Astrophysics. CG was supported by FAPESP (Funda\c{c}\~ao de Amparo \'a Pesquisa do Estado de S\~ao Paulo) under grant 2021/01183-8.
}

\bibliography{sgra_polar}
\bibliographystyle{aa}

\appendix
\section{2017 Apr 7 and 11 data}
\label{app:table_711}

\begin{table*}[h!]
     \caption{Summary of the properties of \sgra light curves observed with ALMA in April 2017
     }
    \begin{center}
    \setlength{\tabcolsep}{3.2pt}
    \renewcommand{\arraystretch}{1.1} 
    \begin{tabularx}{0.995\linewidth}{lcccccccccc}
    \hline
    \hline
      & \multicolumn{4}{c}{ Band 6, 2017 Apr 7}  &  &\multicolumn{4}{c}{ Band 6, 2017 Apr 11}   \\
\hline
      $\nu_{\rm obs}$ (GHz) & 213.1 & 215.1 & 227.1$^a$ & 229.1 & & 213.1 & 215.1 & 227.1$^a$ & 229.1\\
      $\mathcal{I}$ (Jy) & $2.35\pm 0.14$ & $2.35 \pm 0.14$ & $2.35\pm 0.15$ & $2.35\pm 0.15$ & &  $2.36 \pm 0.29$ & $2.34 \pm 0.30$ &  $2.32 \pm 0.29$ &  $2.32 \pm 0.30$ \\
      $|\mathcal{P}|$ (mJy) & $168.9\pm 56.3$ & $170.7\pm57.3$ & $181.4\pm 63.0$ & $181.1\pm 63.6$ & & $190.0\pm 50.5$ & $188.8\pm 50.4$ & $195.7\pm 52.1$ & $199.1\pm 54.9$ \\
      $p$ (\%) & $7.15\pm 2.32$ & $7.23 \pm 2.37$ & $7.65\pm 2.57$ & $7.67\pm 2.62$ & &
      $8.02\pm 1.78$ & $8.01 \pm 1.80 $ & $8.40 \pm 1.98$ & $8.53 \pm 2.05$ \\
      $\chi^b$ (deg) & $111.9 \pm 6.2$ & $112.8 \pm 6.2$ & $118.0\pm 6.9$ & $118.6\pm 7.2$ & &
      $128.8\pm 8.9$ & $129.5\pm 9.1$ & $133.5\pm 9.3$ & $133.9\pm 9.2$ \\
      $\mathcal{V}$ (mJy) & $-27.6\pm 9.8$ & $-27.5 \pm 9.8 $ & $-27.9\pm 9.9$ & $-27.7\pm 10.1$ & &
      $-29.7\pm 5.8$ & $-27.7\pm 6.0 $ & $-26.0 \pm 5.8$ & $-24.7 \pm 5.4$ \\
      $v$ (\%) & $-1.18 \pm 0.43$ & $-1.18 \pm 0.43$ & $-1.19\pm 0.42$ & $-1.18 \pm 0.43$ & &  
      $-1.30 \pm 0.38$ & $-1.22 \pm 0.39$ & $-1.16 \pm 0.37$ & $-1.10 \pm 0.35$ \\
    $\alpha_{\mathcal{I}}$ & \multicolumn{4}{c}{ $0.02 \pm 0.01$} & &  \multicolumn{4}{c}{ $-0.20 \pm 0.01$} \\
    $\alpha_{p}$ & \multicolumn{4}{c}{ $1.02 \pm 0.03$} & &  \multicolumn{4}{c}{ $0.78 \pm 0.02$} \\
      $\alpha_{v}$ & \multicolumn{4}{c}{ $-0.07 \pm 0.03$ } & &  \multicolumn{4}{c}{ $-1.87 \pm 0.03$} \\
      RM$^c$ & \multicolumn{4}{c}{  $-4.50 \pm 1.16 $} & & \multicolumn{4}{c}{$-3.19 \pm 0.72 $} \\
     \hline    
    \hline
    \end{tabularx}
    \label{tab:LC_statistics_cd}
    \end{center}
    $^a$scaled up by 4\% to account for the CN absorption line \citep[Appendix H.1. of][]{Goddi2021};\\
    $^b$calculated using directional statistics;  \ $^c$ in the units of $10^5$ rad m$^{-2}$\\
\end{table*}

We present a summary of the analysis of ALMA light curves obtained on 2017 April 7 and 11, in the same fashion as presented in Tab. \ref{tab:LC_statistics} for the 2017 Apr 3 and 6 data. On April 11 an X-ray flare occurred shortly before the ALMA observations \citep{SgraP2,Wielgus2022_LP}. On this day the 230\,GHz light curves were significantly more variable, with a more negative spectral index and lower RM. These changes are broadly consistent with a hotter, more optically thin accretion disk on April 11 \citep[see also][]{Wielgus2022_LC,Wielgus2022_LP}.


\end{document}